\documentclass[onecolumn]{biophys-new}
\usepackage[utf8]{inputenc}
\usepackage{graphicx}
\usepackage[colorlinks,allcolors=cyan!70!black]{hyperref}
\usepackage{textcomp}
\usepackage{units}
\usepackage{mathtools}

\newcommand{\um}{\text{\textmu{}m}}
\newcommand{\etal}{\textit{et~al.}}

\title{Red blood cell shape transitions and dynamics in time-dependent capillary flows}
\runningtitle{Red blood cell dynamics in time-dependent capillary flows} 

\author[1,a,*]{Steffen M. Recktenwald}
\author[2,a]{Katharina Graessel}
\author[1]{Felix M. Maurer}
\author[1]{Thomas John}
\author[2]{Stephan Gekle}
\author[1,3]{Christian Wagner}
\runningauthor{Recktenwald \etal} 

\affil[1]{Dynamics of Fluids, Department of Experimental Physics, Saarland University, Saarbr{\"u}cken, Germany}
\affil[2]{Biofluid Simulation and Modeling, Department of Physics, University of Bayreuth, Bayreuth, Germany}
\affil[3]{Physics and Materials Science Research Unit, University of Luxembourg, Luxembourg, Luxembourg}

\affil[a]{These authors contributed equally to this work.}
\corrauthor[*]{steffen.recktenwald@uni-saarland.de}
\papertype{Article}

\begin{document}

\begin{frontmatter}

\begin{abstract}
The dynamics of single red blood cells (RBCs) determine microvascular blood flow by adapting their shape to the flow conditions in the narrow vessels. In this study, we explore the dynamics and shape transitions of RBCs on the cellular scale under confined and unsteady flow conditions using a combination of microfluidic experiments and numerical simulations. Tracking RBCs in a comoving frame in time-dependent flows reveals that the mean transition time from the symmetric croissant to the off-centered, non-symmetric slipper shape is significantly faster than the opposite shape transition, which exhibits pronounced cell rotations. Complementary simulations indicate that these dynamics depend on the orientation of the RBC membrane in the channel during the time-dependent flow. Moreover, we show how the tank-treading movement of slipper-shaped RBCs in combination with the narrow channel leads to oscillations of the cell’s center of mass. The frequency of these oscillations depends on the cell velocity, the viscosity of the surrounding fluid, and the cytosol viscosity. These results provide a potential framework to identify and study pathological changes of RBC properties.
\end{abstract}

\end{frontmatter}

\section*{Introduction}
Red blood cells (RBCs) are the main constituent of human blood and deliver oxygen to the body tissues via the cardiovascular system. In the smallest vessels of this extensive network, RBCs flow in single file through microvessels with a diameter comparable to their size. At rest, healthy RBCs have a biconcave, discocyte shape with a diameter of roughly \mbox{$\unit[8]{\um}$} and a thickness of \mbox{$\unit[2]{\um}$}. From a mechanical perspective, RBCs consist of a thin lipid bilayer membrane, with a cytoskeleton network on its inner surface, containing the inner cytosol \cite{Secomb2017}. Due to their high deformability, RBCs can deform into various shapes when flowing through microvessels \cite{Skalak1969, Bagge1980, Pries2008} and even pass through capillaries smaller than their own size \cite{Freund2013, Salehyar2016}. This ability to dynamically adapt their shape to the flow conditions in the circulatory system determines the unique flow properties of blood in the microcirculation and is crucial for microvascular RBC transport. 

To understand the deformation and dynamics of RBCs in the microcirculation, the flow behavior of single RBCs at the micrometer scale has been studied both experimentally \cite{Gaehtgens1980, Suzuki1996, Abkarian2006, Tomaiuolo2009, Tomaiuolo2011, Braunmuller2012a, Cluitmans2014, Prado2015, Guckenberger2018, Saadat2020} as well as numerically \cite{Secomb1986, Noguchi2005, Secomb2007, Kaoui2009, Fedosov2010, Kaoui2011, Tahiri2013, Aouane2014, Fedosov2014a,  Lazaro2014, Fedosov2014, Ye2014, Ye2017, Guckenberger2018, Reichel2019, Dasanna2021}. In general, RBCs flowing in small vessels can exhibit a plethora of shapes depending on the channel confinement, the flow velocity, the viscosity of the surrounding fluid, and the mechanical properties of the RBC itself \cite{Abkarian2008, Guido2009}. In microfluidic experiments, RBC suspensions are commonly studied in round and rectangular capillaries of the same order of diameter as RBCs mainly focusing on the deformation and the resulting shapes of single RBCs under steady flow conditions. Two of the most frequently observed shapes are the parachute and the slipper shape. While nearly axisymmetric parachute-shaped RBCs flow in the channel center, non-axisymmetric slippers flow at an off-centered position closer to the vessel walls. In rectangular channels without the fourfold symmetry, parachute-shaped RBCs become more asymmetric and are therefore often called croissants. Guckenberger~\etal~\cite{Guckenberger2018} investigated these two dominant RBC shapes under steady flow conditions in a rectangular microfluidic channel with a width of \mbox{$\unit [12]{\um}$} and a height of \mbox{$\unit [10]{\um}$} employing microfluidic experiments in combination with numerical simulations. Covering a broad velocity range between \mbox{$\unit[0.1]{mm\,s^{-1}}$} and \mbox{$\unit[10]{mm\,s^{-1}}$}, they constructed a phase diagram for stable RBC shapes at a fixed position along the channel's flow direction. The authors observed that croissants predominantly appear at lower velocities, while slippers are generally found at higher cell velocities. However, numerical simulations have shown that the occurrence of these stable RBC shapes under steady-state conditions is affected by the channel geometry, the initial RBC shape and orientation, and the initial cell position in the channel cross-section at the beginning of the microfluidic channel \cite{Guckenberger2018}. In addition to the stable croissant and slipper shapes, various dynamical states including snaking, tumbling, swinging, and tank-treading motions of RBCs have been reported in numerical \cite{Skotheim2007, Dodson2010, Kaoui2011, Tahiri2013, Aouane2014, Fedosov2014, Mauer2018, Reichel2019, Mignon2021, Dasanna2021} and experimental studies \cite{Fischer1978, Tran-Son-Tay1984, Fischer2007, Fischer2004, Abkarian2007, Dupire2012, Lanotte2016}. One of the most fascinating dynamics is the characteristic tank-treading motion of slipper-shaped RBCs, where the cell maintains an almost constant shape, while its membrane circulates around its interior \cite{Fischer1978, Skotheim2007}. Early rheoscopic experiments in shear flow showed that the tank-treading dynamics depend on the applied shear rate and the viscosity of the surrounding fluid \cite{Fischer1978, Tran-Son-Tay1984, Fischer2007}. Recent numerical simulations in microchannel flow further revealed a strong effect of the confinement and the ratio between the inner viscosity of the cell and the viscosity of the outer fluid on RBC dynamics \cite{Yazdani2011, Fedosov2014, Cordasco2014, Sinha2015, Dasanna2021}. 

Besides these snapshot observations under steady conditions, RBCs can exhibit complex behavior under transient flow conditions when entering small vessels from a larger reservoir and at bifurcations \cite{Abkarian2008, Prado2015, Saadat2020}. During the start-up of capillary flow in microfluidic experiments, RBCs were found to transition from the initial biconcave shape at rest to a final parachute-like configuration \cite{Tomaiuolo2011}. The time scale of this transient shape evolution from a shape at rest to a cell shape in flow was roughly \mbox{$\unit[0.1]{s}$} and the final stable RBC shape depended on the cell velocity, the cell orientation in the channel, and the cell size. 

Nevertheless, detailed transitions between different RBC shapes in flow could not be experimentally studied so far, especially under time-dependent flow conditions. From a technical perspective, video microscopy imaging is nowadays commonly employed at a fixed channel position along the flow direction. However, this limits investigations of RBC dynamics to a fixed field of view of merely a few hundred micrometers, depending on the used camera and magnification. To overcome such limitations, we introduce a microscope-based adaptive tracking technique to characterize the dynamics and shape transitions of RBCs in confined microfluidic channels under arbitrary time-dependent flow conditions. This approach allows us to image the shape of single RBCs and extract the complete trajectory of the cell along the whole channel length of several centimeters with cell velocities up to \mbox{$\unit[10]{mm\,s^{-1}}$}, similar to the range of RBC velocities found in the vessels of the microcirculation \cite{Pries1995, Secomb2017}. Applying time-dependent pressure modulations in combination with our tracking method, we show that the transition between the croissant to slipper shape is significantly faster than vice versa and that these transitions are not affected by the applied pressure amplitude. Complementary three-dimensional (3D) numerical simulations indicate the sensitivity of the system to the exact orientation of the cell during and even before the transition process. Additionally, our method enables us to experimentally study the dynamics of slipper-shaped RBCs that show tank-treading motion in confined microscale flows. The observed oscillation amplitudes are in quantitative agreement with corresponding simulations. 

\section*{Materials and Methods}

\subsection*{Experimental}

\subsubsection*{Preparation of red blood cell suspensions}
Blood is taken with informed consent from healthy voluntary donors by needle-prick and suspended in phosphate-buffered saline solution (Gibco PBS, Fisher Scientific; Schwerte, Germany). The suspension is centrifuged at \mbox{$\unit[1500]{g}$} for five minutes. Subsequently, sedimented RBCs are resuspended in PBS and the centrifugation and washing steps are repeated three times. Finally, a hematocrit concentration of \mbox{$\unit [0.5]{\%Ht}$} is adjusted in a PBS solution that contains \mbox{$\unit [1]{g\,L^{-1}}$} bovine serum albumin (BSA, Sigma-Aldrich; Taufkirchen, Germany). Blood withdrawal, sample preparation, and experiments were performed according to regulations and protocols that were approved by the ethics committee of the `Aerztekammer des Saarlandes' (reference No 24/12).

Further, RBCs are suspended in dextran solutions (Dextran \mbox{$\unit [70]{kDa}$}, Sigma-Aldrich; Taufkirchen, Germany) of various concentrations to study the effect of the viscosity of the surrounding fluid on RBC dynamics. Therefore, dextran solutions are prepared in the PBS/BSA mixture and the steady shear viscosity of the solutions without RBCs is determined using a rotational rheometer (MCR 702, Anton Paar; Graz, Austria), equipped with a coaxial cylinder fixture (CC20, inner radius \mbox{$R_{\mathrm{i}}=\unit [10]{mm}$}, outer radius \mbox{$R_{\mathrm{o}}=\unit [11]{mm}$}). The shear viscosity and density values of the dextran solutions are listed in Table~S1 in the Supporting Material.

\subsubsection*{Microfluidic setup}
\begin{figure}[hbt!]
\centering
\includegraphics[width=8.255cm]{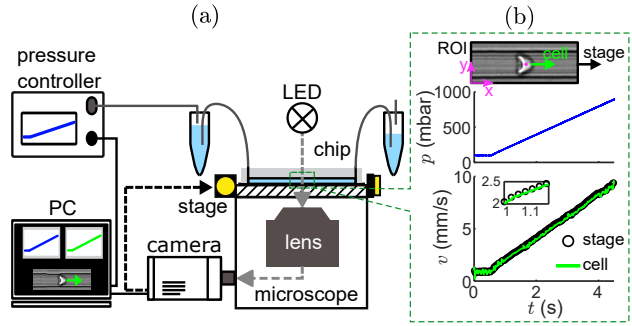}
\caption{(a) Schematic of the experimental setup. (b) Tracking of a RBC during an increasing pressure drop ramp from \mbox{$\unit[100]{mbar}$} to \mbox{$\unit[1000]{mbar}$}. While the cell travels through the channel, as shown in the snapshot, the stage speed is adjusted in a feedback loop to keep the cell in the field of view. The velocity of the cell (green line) is calculated for each frame based on the stage velocity (black circles) and the position of the cell in the field of view (magenta coordinates).}
\label{Figure_Setup}
\end{figure}

The RBC suspensions are pumped through microfluidic channels with a high-precision pressure device (OB1-MK3, Elveflow; Paris, France) that operates in a pressure range of \mbox{$0\leq p\leq \unit[2000]{mbar}$}. The microfluidic device is fabricated using polydimethylsiloxane (PDMS, RTV 615A/B, Momentive Performance Materials; Waterford, New York) through standard soft lithography \cite{Friend2010}. The microfluidic chip consists of 30 parallel channels, each with a rectangular cross-section of width \mbox{$W=\unit[10.8\pm 0.6]{\um}$} in $y$-direction and height \mbox{$H=\unit[7.9\pm 0.3]{\um}$} in $z$-direction. The total length of the channels is \mbox{$L\approx\unit[40]{mm}$}. The channels are connected at the beginning and end where they share common fluid inlet and outlet reservoirs. The inlet and outlet are connected with rigid but flexible medical-grade polyethylene tubing (\mbox{$\unit[0.86]{mm}$} inner diameter, Scientific Commodities; Lake Havasu City, Arizona) to the sample and waste container, respectively. The microfluidic device is mounted on an inverted microscope (Eclipse TE2000-S, Nikon; Melville, New York), equipped with a computer-controlled, motorized stage (Scanning Stage SCAN IM 120 x 100, M{\"a}rzh{\"a}user Wetzlar; Wetzlar, Germany), a USB~3.0 camera (DMK 23U1300, The Imaging Source; Bremen, Germany), and a LED illumination (Fig.~\ref{Figure_Setup}(a)). For the cell tracking experiments, a \mbox{$20\times$} air objective (Plan Fluor, Nikon; Melville, USA) with a numerical aperture \mbox{$NA=0.45$} is used. 

To determine the fraction of RBC shapes in the channel under steady flow conditions, shapes are classified at a constant pressure drop covering a broad pressure range of \mbox{$20 \leq p \leq \unit[1000]{mbar}$}. Therefore, RBCs are recorded with a frame rate of \mbox{$\unit[400]{fps}$} at a fixed position at the middle of the microfluidic chip at \mbox{$x=L/2$} using a high-magnification \mbox{$60\times$} oil immersion objective with a high \mbox{$NA=1.55$} (CFI Plan Fluor, Nikon; Melville, New York). RBC shapes are classified manually according to Guckenberger~\etal \cite{Guckenberger2018} and the individual cell velocity is determined by tracking the cell position over the image sequence within the field of view.

\subsubsection*{RBC tracking technique}
In this study, we present a microfluidic technique that allows us to track objects, \textit{e.g.}, RBCs, in a comoving frame over several centimeters along the channel flow direction. To obtain the comoving frame, we use a customized feedback-control mechanism between the computer-controlled, motorized stage and the image stream from the camera, implemented in a self-written MATLAB program (The MathWorks; Natick, Massachusetts). Initially, the ROI of the camera, which operates at a frame rate of \mbox{$\unit[100]{fps}$}, is positioned stationary close to the channel entrance without any passing cells. The positions of the channel borders are detected and the image background is calculated by computing a pixel-wise average over multiple frames. When a cell enters the region of interest (ROI), the number of pixels above the background level exceeds an adjustable threshold value that is related to the cell projection area and the tracking routine loop starts. During the tracking loop, the position of the cell is determined in every frame and a consistency check is performed based on the position of the cell in the previous frame. Every second frame, the stage velocity is adjusted using a non-linear integral controller to keep the cell at a given position in the field of view. This procedure is repeated until the cell exits the channel. To keep the traffic of the serial communication interface low while using a frame rate of \mbox{$\unit[100]{fps}$}, the stage position and velocity are saved every twelfth frame during the feedback loop. As schematically shown in Fig.~\ref{Figure_Setup}(b) for a pressure ramping from \mbox{$\unit[100]{mbar}$} to \mbox{$\unit[1000]{mbar}$} (blue line), the RBC velocity (green line) is calculated for each frame based on the stage velocity (black circles) and the position of the RBC in each frame. The maximum cell velocity that can be tracked is approximately \mbox{$\unit[10]{mm\,s^{-1}}$}, limited by the speed of the motorized stage. All tracking experiments are performed at room temperature.

\subsection*{Simulations}
We perform numerical simulations of single RBCs in a rectangular channel that closely matches the experimental settings. The channel has a width of \mbox{$W=\unit[11.5]{\um}$} and a height of \mbox{$H=\unit[8]{\um}$} with slightly rounded corners. The length of the channel is \mbox{$L=\unit[42.7]{\um}$}, and periodic boundary conditions are applied in the flow direction.

The Reynolds number \mbox{$\text{Re}$}, which relates the inertial to viscous forces in the system, is defined as \mbox{$\text{Re} = {v D_{\mathrm{h}}\rho}/{\eta_\mathrm{o}}$}, with the fluid density \mbox{$\rho=\unit[1]{g\,cm^{-3}}$}, the cell velocity \mbox{$v$}, the fluid's dynamic viscosity \mbox{$\eta_\mathrm{o}=\unit[1.2]{mPa\,s}$}, and the hydraulic diameter of the rectangular microfluidic channel \mbox{$D_{\mathrm{h}}=2W H/(W+H)$}. For characteristic velocities \mbox{$v<\unit[10]{mm\,s^{-1}}$}, the maximum Reynolds number is \mbox{$\text{Re}<0.1$}, hence the Stokes equation adequately describes the flow. Thus, we use the boundary integral method to compute the trajectory and deformation of the cell moving  through the channel \cite{Pozrikidis2001, Zhao2010, Guckenberger2018, Guckenberger2018a}.

The RBC has a long radius of \mbox{$R=\unit[3.91]{\um}$} when in the characteristic discocyte shape, a surface area of \mbox{$A=\unit[133.5]{\um^2}$}, and a volume of \mbox{$V=\unit[93.5]{\um^3}$} \cite{Evans1972, Skalak1989, Diez-Silva2010}. Its surface is discretized by 2048 flat triangles. The infinitely thin elastic membrane of the RBC is endowed with resistance to shear and area dilation and is modeled using Skalak's law \cite{Skalak1973, Barthes-Biesel2002} with the shear modulus \mbox{$\kappa_\mathrm{S}=\unit[5 \times 10^{-6}]{N\,m^{-1}}$} \cite{Mills2004, Yoon2008, Freund2014} and the area dilation modulus \mbox{$\kappa_\mathrm{A}=100\kappa_\mathrm{S}$}. The typical discocyte shape is used as an elastic reference shape for the shear contribution unless stated otherwise. An alternative choice for the elastic reference shape is an oblate spheroid \cite{Viallat2014, Sinha2015, Cordasco2014, Peng2014, Levant2016} with the same surface area as the RBC. Some simulations are additionally performed with oblate spheroids of aspect ratio \mbox{$\tau=0.9$} or \mbox{$\tau=0.98$} as an elastic reference shape. The RBC membrane shows resistance to bending implemented using the Helfrich law \cite{Canham1970, Helfrich1973, Guckenberger2016, Guckenberger2017}  with a bending modulus of \mbox{$\kappa_\mathrm{B}=\unit[3\times 10^{-19}]{N\,m}$} \cite{Evans1983, Freund2014} and a flat bending reference shape. 

The Newtonian fluid flowing through the channel has a dynamic viscosity \mbox{$\eta_\mathrm{o}$}. The RBC is filled with another Newtonian fluid with a viscosity \mbox{$\eta_\mathrm{i}$}, resulting in a viscosity contrast \mbox{$\lambda=\eta_\mathrm{i}/\eta_\mathrm{o}$}. Early experimental investigations suggested that the physiological viscosity contrast is roughly \mbox{$\lambda=5$} \cite{Cokelet1968, Wells1969}. Hence, numerical simulations of RBCs are often performed with \mbox{$\lambda=5$} \cite{Guckenberger2018, Reichel2019}. However, we found that simulation results for \mbox{$\lambda=10$} are closer to the experimental observations. Therefore, simulations are performed at \mbox{$\lambda=10$} in this study unless stated otherwise. Choosing a higher viscosity contrast might also provide a potential way to account for the viscosity of the membrane \cite{Yazdani2013, Fischer2015}, which is not explicitly included in our numerical model.

In the simulations, we do not apply a pressure gradient but control the RBC velocity by setting a mean flow velocity. To mimic the experimental pressure ramps, the cell accelerates linearly over time from a lower \mbox{$v_{\mathrm{low}}$} to a higher \mbox{$v_{\mathrm{high}}$} mean velocity. Simulations are performed up to cell velocities of \mbox{$\unit[8]{mm\,s^{-1}}$}, since they become unstable beyond that point due to the high shear rates in the narrow channel. The velocity ramp simulations all start with a RBC in the discocyte shape centered in the channel with its symmetry axis aligned with the channel axis. Before the start of the ramp, the cells flowed for a short time of \mbox{$\unit[0.2]{s}$} at the lower velocity \mbox{$v_{\mathrm{low}}$}. This reflects the fact that experimental measurements started close to the channel entrance. 

\section*{Results and Discussion}

\subsection*{Stable red blood cell shapes under steady flow conditions}

\begin{figure}[hbt!]
\centering
\includegraphics[width=8.255cm]{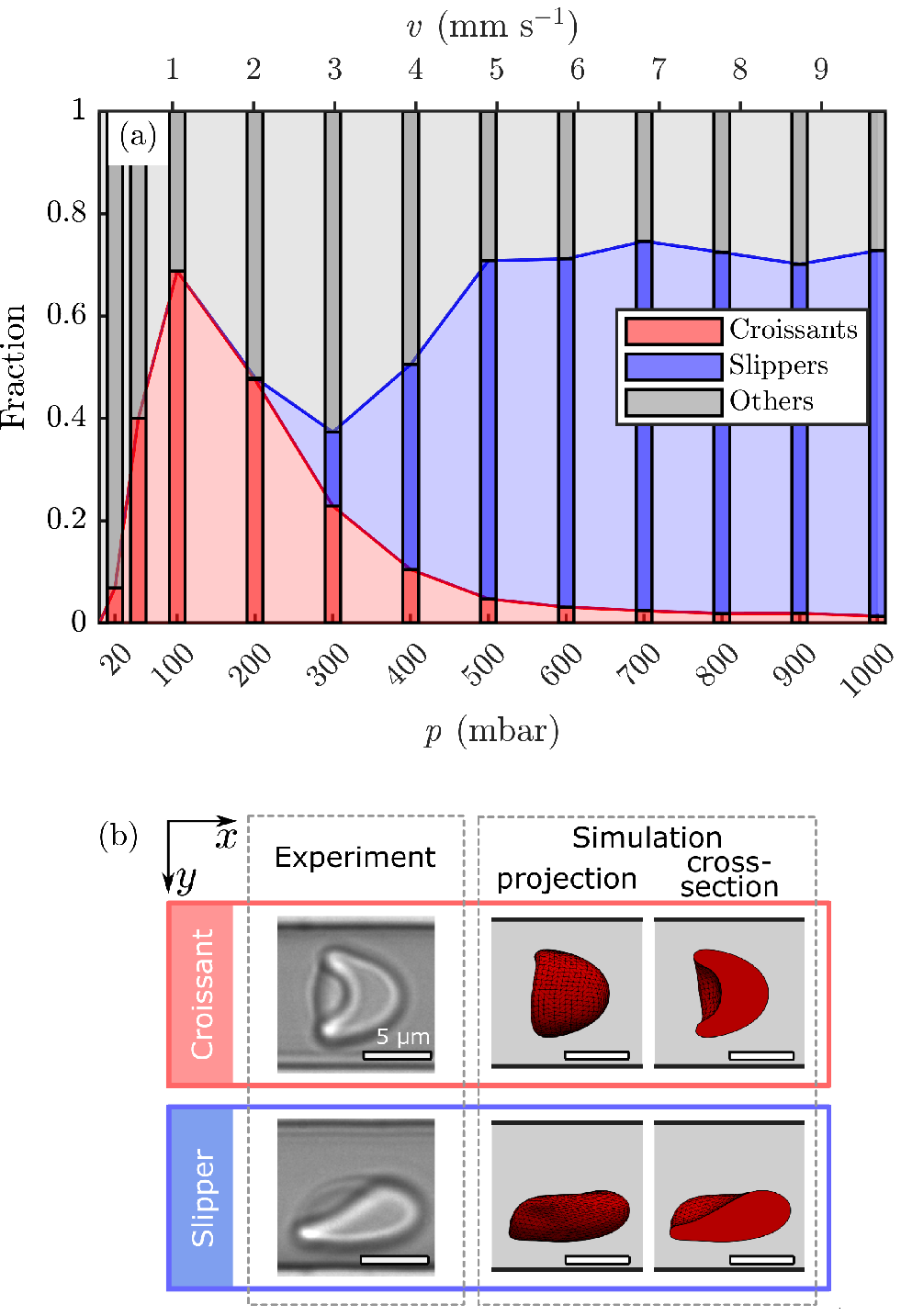}
\caption{Classification of RBC shapes in microfluidic channels under constant flow conditions. (a) Fraction of RBC shapes as a function of the applied pressure drop (bottom axis) and the mean cell velocity (top axis). (b) Representative examples of a croissant (top) and a slipper (bottom) for experiments with \mbox{$p=\unit[100]{mbar}$} and \mbox{$p=\unit[600]{mbar}$}, corresponding to cell velocities of \mbox{$v\approx\unit[1]{mm\,s^{-1}}$} and \mbox{$v\approx\unit[5.8]{mm\,s^{-1}}$}, respectively. The simulation snapshots are obtained at similar velocities. The flow is in $x$-direction and the scale bars represent a length of \mbox{$\unit[5]{\um}$}. }
\label{Figure_PD}
\end{figure}

Depending on their velocity and the channel confinement RBCs exhibit various shapes when flowing through microchannels. Here, cells are classified under steady flow conditions either as one of the two most dominant RBC shapes, \textit{i.e.}, croissants and slippers, or as others that are not categorized further \cite{Guckenberger2018}.
To derive the shape phase diagram, measurements are taken at a fixed position at \mbox{$x=L/2$}, and cells are classified based on their shape in this fixed field of view. We found with our adaptive control setup that individual RBCs keep their stable shapes while traveling through at channel under steady flow conditions. The fractions of cells in each of the three main classes at a constant pressure drop  in the microfluidic experiments are plotted in Fig.~\ref{Figure_PD}(a). The corresponding mean cell velocities are shown at the top $x$-axis, covering a range of \mbox{$0.1 \leq v \leq \unit[9.7]{mm\,s^{-1}}$}, similar to RBC velocities found in the microcirculation \cite{Pries1995, Secomb2017}. The centered croissants appear predominantly at lower pressure drops with a pronounced peak of roughly \mbox{$70\%$} at \mbox{$p=\unit[100]{mbar}$} or \mbox{$v\approx\unit[1]{mm\,s^{-1}}$}. For pressure drops above \mbox{$p \geq \unit[300]{mbar}$} (\mbox{$v\approx\unit[3]{mm\,s^{-1}}$}), the fraction of croissants significantly decreases while off-centered slipper-shaped RBCs emerge. Above \mbox{$p \geq \unit[500]{mbar}$} (\mbox{$v\approx\unit[5]{mm\,s^{-1}}$}), croissants disappear almost completely and the fraction of slippers plateaus at roughly \mbox{$70\%$}. In the numerical simulations, croissant and slipper shapes appear at velocities similar to those of the experiments, as shown in Fig.~S1 in the Supporting Material. Representative images of a croissant and a slipper-shaped RBC are shown in Fig.~\ref{Figure_PD}(b). The phase diagrams in Fig.~\ref{Figure_PD} and Fig.~S1 are in good agreement with studies in similar microchannels \cite{Guckenberger2018} and serve as a reference point to define the pressure and velocity limits of the time-dependent flow modulations. 

\subsection*{Red blood cell shape transitions in time-dependent flows}

\begin{figure*}[hbt!]
\centering
\includegraphics[width=17.145cm]{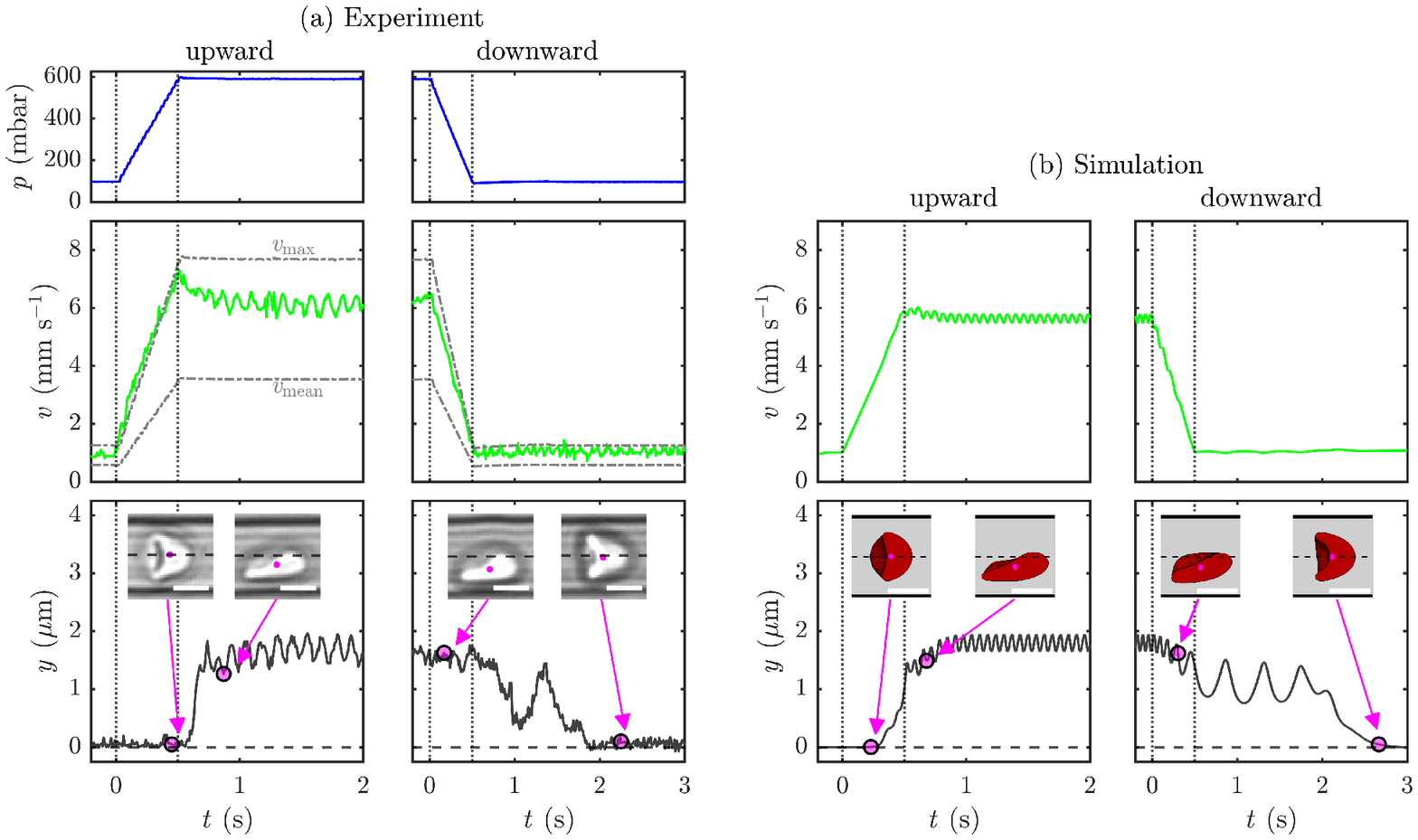}
\caption{Dynamics of single RBCs in time-dependent flows at \mbox{$t_{\mathrm{ramp}}=\unit[0.5]{s}$} for (a) an experiment and (b) a simulation. The left and right columns in (a) and (b) correspond to upward and downward ramps, respectively. In (a), the top panels show the applied pressure drop signal over time, ramping from \mbox{$p_{\mathrm{low}}=\unit[100]{mbar}$} to \mbox{$p_{\mathrm{high}}=\unit[600]{mbar}$} for the upward ramp, and vice versa for the downward situation. The middle panels in (a) show the cell velocity as green line. The gray dashed-dotted lines correspond to analytical solutions of the maximum and mean velocity inside the channel. The corresponding simulations in (b) show the cells transitioning from a velocity \mbox{$v_{\mathrm{low}}\approx\unit[1]{mm\,s^{-1}}$} to \mbox{$v_{\mathrm{high}}\approx\unit[5.8]{mm\,s^{-1}}$}. Vertical dotted lines indicate the start and end of the ramps. The bottom panels in (a) and (b) show the $y$-coordinate of the cell's center of mass. The inset images show the cells at the start and end points of the shape transitions, highlighted by the magenta markers. The horizontal, dashed black lines correspond to the channel center axis. The scale bars represent a length of \mbox{$\unit[5]{\um}$}. Corresponding videos are provided in the Supporting Material for the upward experiment and simulation (Movie~S1 and Movie~S2) and for the downward experiment and simulation (Movie~S3 and Movie~S4).}
\label{Figure_Ramp}
\end{figure*}

To study the dynamical transitions between the stable croissant and slipper shape experimentally, we apply pressure drop ramps between a low \mbox{$p_{\mathrm{low}}$} and a high \mbox{$p_{\mathrm{high}}$} pressure limit within a ramping time \mbox{$t_{\mathrm{ramp}}$}. The pressure limits are chosen such that at \mbox{$p_{\mathrm{low}}=\unit[100]{mbar}$} RBCs mainly exhibit the croissant shape, while in the range of \mbox{$600 \leq p_{\mathrm{high}}\leq \unit[900]{mbar}$} slippers appear predominantly. The total pressure drop change during the ramp is consequently defined as \mbox{$\Delta p = p_{\mathrm{high}}-p_{\mathrm{low}}$}. These ramping experiments are conducted in upward and downward ramp direction with ramping times \mbox{$t_{\mathrm{ramp}}$} covering a range between \mbox{$\unit[0.125]{s}$} and \mbox{$\unit[1]{s}$}.

Representative results for ramps with \mbox{$t_{\mathrm{ramp}}=\unit[0.5]{s}$} are shown in Fig.~\ref{Figure_Ramp}(a) and (b) for an experiment and a corresponding simulation, respectively. In Fig.~\ref{Figure_Ramp}(a), the applied pressure limits are \mbox{$p_{\mathrm{low}}=\unit[100]{mbar}$} and  \mbox{$p_{\mathrm{high}}=\unit[600]{mbar}$}, resulting in similar RBC velocities as the mean velocities of \mbox{$v_{\mathrm{low}}\approx\unit[1]{mm\,s^{-1}}$} and \mbox{$v_{\mathrm{high}}\approx\unit[5.8]{mm\,s^{-1}}$} in the numerical simulation in Fig.~\ref{Figure_Ramp}(b). The origin of the time \mbox{$t=0$} is set to the beginning of the ramps. As the applied pressure changes over time in the experiment, the RBC velocity changes without any time delay due to the negligible inertia in the system \mbox{$\text{Re}\approx0.05$}.

To characterize the RBC shape transition, the $y$-coordinate of the cell's center of mass is plotted as a function of time in the bottom row of Fig.~\ref{Figure_Ramp}. Corresponding videos of the cells' movement in the comoving frame are provided in Movie~S1 and Movie~S2 in the Supporting Material for the upward experiment and simulation, respectively. During the upward ramp, the RBC initially flows as a croissant at the channel centerline at \mbox{$y=0$}. As the velocity increases during the upward ramp, the cell stays in the channel middle. Eventually, the RBC directly transitions from the croissant into the off-centered slipper shape, resulting in a change of the $y$-coordinate of the cell's center of mass, similar to the upward numerical simulation in Fig.~\ref{Figure_Ramp}(b). The beginning and end of the shape transition are highlighted with magenta markers in the bottom row of Fig.~\ref{Figure_Ramp}. They are determined based on the pictures of the cell shape, analogously to the manual shape classification used for the construction of the phase diagram in Fig.~\ref{Figure_PD}. Although the pressure or velocity ramp starts at \mbox{$t=0$}, the beginning of the transition process is delayed for the upward ramp and begins at \mbox{$t\approx\unit[0.4]{s}$} and \mbox{$t\approx\unit[0.23]{s}$} for the experiment and simulation, respectively. The shape transition is accompanied by a shift of the cell's center of mass from the channel center to a position closer to one of the channel's side faces in $y$-direction. Note that this transition occurs randomly to one of the channel side walls in positive or negative $y$-direction. This movement to a streamline closer to the wall results in a slight decrease of the RBC velocity after reaching the end of the pressure ramp at \mbox{$t=\unit[0.5]{s}$}, which is also seen in the simulation. Once the RBC fully transitioned into the slipper shape and reaches the off-centered steady-state position, we observe pronounced oscillations of the cell's center of mass over time in both the experiment and the simulation. These dynamics lead to a continuous cross-streamwise oscillation of the slipper-shaped RBC while flowing through the channel, which results in an oscillating velocity. 

Similar to the upward ramp experiment, the right panels in Fig.~\ref{Figure_Ramp}(a) and (b) show the downward ramp experiment and simulation, respectively. Movie~S3 and Movie~S4 in the Supporting Material show corresponding videos of the cells' movement in the comoving frame for the downward experiment and simulation, respectively. In the simulations, the downward ramp starts after a plateau at \mbox{$v_\mathrm{high}$}, which lasts for \mbox{$t\approx\unit[1.5]{s}$}. In the downward scenario, the cell initially is in the off-centered slipper shape. During the decrease of the applied pressure or velocity, the shape transitions start with a time delay at \mbox{$t\approx\unit[0.18]{s}$} in the experiment and \mbox{$t\approx\unit[0.3]{s}$} in the simulation with respect to the start of the downward ramp at \mbox{$t=0$}. However, in contrast to the direct shape transition during the upward ramp, the cell shown for the downward case in Fig.~\ref{Figure_Ramp}(a) does not immediately transition from one shape into the other, but exhibits a pronounced cell rotation, which is accompanied by a deformation of the membrane and a tumbling motion of the cell. For this downward experiment, the RBC first approaches the channel middle, then rotates and tumbles once, as indicated by the increase in the $y$-coordinate at \mbox{$t\approx\unit[1.3]{s}$}, before finally achieving its steady-state position in the channel center. Multiple similar cell rotations during the shape transition also occur during the downward simulation in Fig.~\ref{Figure_Ramp}(b). To differentiate such shape transitions from the typically direct shape transitions during the upward ramps, we refer to RBC shape changes that exhibit pronounced rotations as rotating shape transitions. While \mbox{$29\%$} of all investigated RBCs show pronounced rotating transitions during the downward ramping experiments, such transitions are found in merely \mbox{$5\%$} of all upward ramping experiments. Additional examples for direct and rotating transitions for experiments and simulations are provided in Fig.~S2 in the Supporting Material. 

\begin{figure*}[hbt!]
\centering
\includegraphics[width=17.145cm]{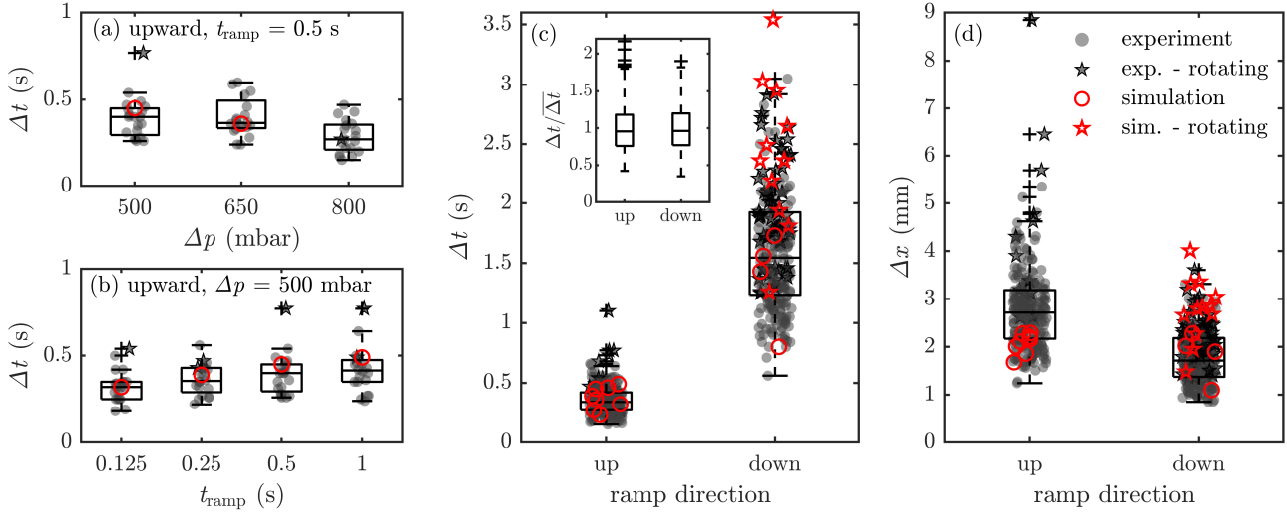}
\caption{Effect of the ramping parameters on the shape transition. Experimental data are shown as boxplots with superimposed individual gray data points. Outliers are plotted using '$+$'-symbols. Simulation results are plotted as red symbols. Cells that show pronounced rotations during the shape transitions are highlighted as stars. (a) Transition time \mbox{$\Delta t$} for an upward ramp as a function of the total pressure difference at a constant ramp duration of \mbox{$t_{\mathrm{ramp}}=\unit[0.5]{s}$}. For \mbox{$\Delta p=\unit[500]{mbar}$} and \mbox{$\Delta p=\unit[650]{mbar}$}, the velocities in the simulations are \mbox{$\Delta v \approx \unit[4.7]{mm\,s^{-1}}$} and \mbox{$\Delta v \approx \unit[6.2]{mm\,s^{-1}}$}, respectively. (b) Transition time for different ramping duration for an upward ramp with \mbox{$\Delta p=\unit[500]{mbar}$}, corresponding to \mbox{$\Delta v \approx \unit[4.7]{mm\,s^{-1}}$} in the numerical simulation. In (c), all individual measurements for the upward and downward ramp directions are combined. The inset shows the experimentally observed transition times normalized by the mean transition time \mbox{$ \overline{\Delta t}$}. (d) Transition length \mbox{$\Delta x$} as a function of the ramping direction, combining all individual measurements, similar to (c). The individual measurements of \mbox{$\Delta t$} and \mbox{$\Delta x$} are shown Fig.~S5 and Fig.~S6 in the Supporting Material, respectively.}
\label{Figure_Trans}
\end{figure*}

As shown in Fig.~\ref{Figure_Ramp}, the shape transitions begin after a certain time delay during the upward and downward ramps. In an effort to evaluate the critical conditions for these shape changes to start, we calculate the transition velocity \mbox{$v_{\mathrm{t}}$} at which the shape transitions begin. For this, the cell velocity at the start of the transition from the croissant to the slipper case is determined during the upward ramps and for the vice-versa transition during the downward ramps. Figure~S3 in the Supporting Material shows \mbox{$v_{\mathrm{t}}$} for various ramping times for (a) the upward and (b) the downward ramp. We include all pressure amplitudes \mbox{$\Delta p$} in Fig.~S3, since we do not observe a correlation between the pressure amplitude and the transition velocity. For the experiments, the transition velocity scatters broadly for both the upward and downward direction, as shown by the black boxplots with superimposed gray data points. Nevertheless, a slightly decreasing mean transition velocity with increasing ramping time \mbox{$t_{\mathrm{ramp}}$} can be observed for both ramping directions. The numerical simulations reveal similar transition velocities for the upward and downward ramp direction and a similar dependency on \mbox{$t_{\mathrm{ramp}}$}. In the case of the downward ramp, the mean experimental values are in good agreement with the numerical predictions, while for the upward ramp direction, experimentally determined \mbox{$v_{\mathrm{t}}$} are overall higher than the numerical predictions. However, this deviation is attributed to a pressure-induced deformation of the PDMS-based microfluidic chip during the upward pressure ramp, where the widening of the channels increases with increasing applied pressure \cite{Recktenwald2021}. This increase of the channel cross-section leads to a higher RBC velocity at a fixed pressure drop, which is also visible in the upward experiment in Fig.~\ref{Figure_Ramp}(a), where the cell velocity (green line) shows a steeper increase than the analytical prediction (gray dashed-dotted lines). This effect also results in velocity overshoots during upward experiments with very short ramping duration, as shown in Fig.~S4 in the Supporting Material. Hence, the beginning of the shape transitions in the upward experiments are often close to this velocity overshoot, which yields higher transition velocities for the experimental upward case in Fig.~S3 with respect to the numerical simulations. In contrast to the upward ramps, the pressure-induced change of the channel cross-section during the downward ramp is less pronounced and therefore we find a better agreement between the experimental and numerical results for the downward case in Fig.~S3. Note that the mean values of the transition velocity \mbox{$2 < v_{\mathrm{t}} < \unit[7]{mm\,s^{-1}}$} lie in the bistability region of the phase diagram, where both croissants and slippers appear \cite{Guckenberger2018}.

To further study the influence of the ramping parameters, \textit{i.e.}, the total pressure amplitude \mbox{$\Delta p$} during the ramp, the ramping duration \mbox{$t_{\mathrm{ramp}}$}, and the ramping direction, on the shape transition, we determine the transition time \mbox{$\Delta t$} that is needed for the shape transitions process. Hence, \mbox{$\Delta t$} is the time duration, which RBCs need to switch from the croissant to the slipper shape in upward ramping experiments and vice-versa in downward measurements, as exemplified in Fig.~\ref{Figure_Ramp}. This transition time is determined based on the cell images from the experiments and simulations, where the beginning corresponds to the point in time when the transition starts at \mbox{$v_{\mathrm{t}}$}. Figure~\ref{Figure_Trans} shows the experimental and simulation results and stars indicate RBCs that show pronounced rotations during the shape transition. As shown in Fig.~\ref{Figure_Trans}(a) for a representative upward ramp at \mbox{$t_{\mathrm{ramp}}=\unit[0.5]{s}$}, the transition time \mbox{$\Delta t$} does not depend on the total pressure amplitude \mbox{$\Delta p$} during the ramp. Surprisingly, we do not find a pronounced effect of the ramping duration \mbox{$t_{\mathrm{ramp}}$} on \mbox{$\Delta t$}, as exemplified in Fig.~\ref{Figure_Trans}(b) for a constant pressure amplitude of \mbox{$\Delta p=\unit[500]{mbar}$}, corresponding to \mbox{$\Delta v \approx \unit[4.7]{mm\,s^{-1}}$} in the numerical simulation. Although, the mean values of \mbox{$\Delta t$} in Fig.~\ref{Figure_Trans}(b) show a slight increase with \mbox{$t_{\mathrm{ramp}}$}, this tendency is not found for other pressure differences, as shown in Fig.~S5 in the Supporting Material. Similar to the observations for an upward ramp, the transition time in downward ramping experiments also does neither depend on \mbox{$t_{\mathrm{ramp}}$} nor \mbox{$\Delta p$} or \mbox{$\Delta v$}. The individual measurements for all combinations of \mbox{$t_{\mathrm{ramp}}$} and \mbox{$\Delta p$} or \mbox{$\Delta v$} are summarized in Fig.~S5 in the Supporting Material. 

However, we observe a striking difference in the transition time depending on the ramping direction, as shown in Fig.~\ref{Figure_Trans}(c). Here, the mean transition time from the croissant to the slipper shape in upward ramp experiments is \mbox{$ \overline{\Delta t}_{\mathrm{up}}=\unit[0.36]{s}$}. In contrast, we observe a much longer transition time of \mbox{$ \overline{\Delta t}_{\mathrm{down}}=\unit[1.6]{s}$} for the opposite shape transition in downward ramp experiments, including a broad scattering of \mbox{$\Delta t$}. For the simulations, we find mean transition times of \mbox{$ \overline{\Delta t}_{\mathrm{up}}=\unit[0.37]{s}$} and \mbox{$ \overline{\Delta t}_{\mathrm{down}}=\unit[2.15]{s}$} for the upward and downward scenarios, respectively. Normalizing the experimental data by the mean transition times \mbox{$ \overline{\Delta t}$}, as depicted in the inset of Fig.~\ref{Figure_Trans}(c), shows that the relative deviations in both ramp directions are very similar. Additionally, cells that rotate and tumble during the shape transitions often exhibit comparatively longer transition times in both ramp directions, indicated by the star symbols in Fig.~\ref{Figure_Trans}(c). 

The difference in the transition time between the two scenarios arises mainly due to two effects. First, the shape transition during the upward ramp starts when the cell reaches the transition velocity \mbox{$v_{\mathrm{t}}$}. This happens either during the ramp or after reaching the plateau at \mbox{$p_{\mathrm{high}}$}, as shown in Fig.~\ref{Figure_Ramp} and Fig.~S4. Hence, the average velocity during the consecutively occurring transition process is close to the plateau velocity at \mbox{$v_{\mathrm{high}}$}. For the upward ramp scenario, this high cell velocity enhances the cross-stream movement of the cell in $y$-direction resulting in the formation and stabilization of the slipper-shaped RBC. In contrast, during the opposite ramp, the transition also starts during the ramp, but most of the process occurs at the lower plateau \mbox{$p_{\mathrm{low}}$}, as shown in Fig.~\ref{Figure_Ramp} and Fig.~S4. Note that the amount of time during \mbox{$\Delta t$} that the cell flows at \mbox{$p_{\mathrm{low}}$} or \mbox{$v_{\mathrm{low}}$} is much larger than the time during the ramp after having reached \mbox{$v_{\mathrm{t}}$}. Thus, the average velocity during the downward transition is close to \mbox{$v_{\mathrm{low}}$}, which leads to a slower transition process from the slipper to the croissant shape and thus a slower movement towards the channel centerline. Guckenberger~\etal~\cite{Guckenberger2018} showed in their simulations that cells at lower velocities and cells which become croissants need more time until they reach the steady-state compared to cells at larger velocities and transitioning into the slipper shape. The second effect is that \mbox{$29\%$} of the RBCs rotate during the downward transition before reaching the stable croissant shape. This rotation at lower cell velocities increases the transition time for the downward ramping experiments. In contrast, only \mbox{$5\%$} of the cells show such rotating shape transitions during the upward ramp, where the shape transition is already accelerated due to the higher cell velocity. 

Further, our tracking approach allows us to determine the transition length \mbox{$\Delta x$} that the RBC travels through the microchannel during the shape transition process. The transition length is calculated by \mbox{${\Delta x}=\int_{0}^{\Delta t} v(t) \,dt$}. Here, \mbox{$t=0$} represents the start of the transition process with \mbox{$v(t=0)=v_{\mathrm{t}}$}. Figure~\ref{Figure_Trans}(d) shows \mbox{$\Delta x$} for both ramp directions, including all combinations of \mbox{$t_{\mathrm{ramp}}$} and \mbox{$\Delta p$} or \mbox{$\Delta v$}. The individual measurements are summarized in Fig.~S6 in the Supporting Material. In case of the upward ramping experiments, we find a slightly higher mean transition length of \mbox{$\overline{\Delta x}_{\mathrm{up}}=\unit[2.8]{mm}$} compared to the downward scenario \mbox{$ \overline{\Delta x}_{\mathrm{down}}=\unit[1.8]{mm}$}. However, the upward data also shows a broader scattering, which is again attributed to the channel widening and the velocity overshoot during the upward ramp, similar to the results discussed in Fig.~3 and Fig.~S4. For the simulations, we observe mean transition lengths of \mbox{$ \overline{\Delta x}_{\mathrm{up}}=\unit[2.1]{mm}$} and \mbox{$ \overline{\Delta x}_{\mathrm{down}}=\unit[2.5]{mm}$} for the upward and downward scenarios, respectively. Assuming an average transition length of \mbox{$\unit[2-3]{mm}$} and a characteristic cell diameter of \mbox{$D\approx\unit[8]{\um}$}, the shape transitions occur over approximately 250 to 375 times the RBC diameter while traveling through the microvessel.

\begin{figure}[hbt!]
\centering
\includegraphics[width=8.255cm]{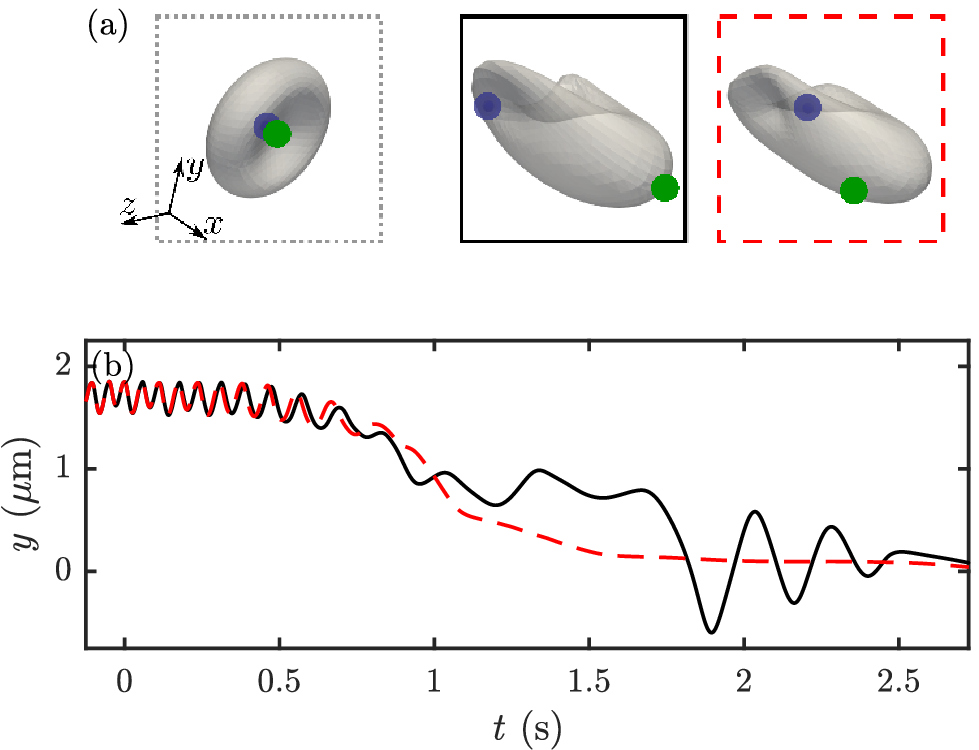}
\caption{Simulated dependency of the downward transition process on the RBC's dimple position for a downward ramp from \mbox{$v_{\mathrm{high}}\approx \unit[7.2]{mm\,s^{-1}}$} to \mbox{$v_{\mathrm{low}}\approx \unit[1]{mm\,s^{-1}}$} within \mbox{$t_{\mathrm{ramp}}=\unit[1]{s}$}. (a) Simulation snapshots of the RBC at rest (left) and at the two starting positions (middle and right) of the downward ramp at \mbox{$t=0$}. (b) $y$-coordinate of the cell's center of mass, corresponding to the two scenarios marked by the black and red boxes in (a). }
\label{Figure_Simu}
\end{figure}

Moreover, our numerical simulations suggest that the transition dynamics during the downward ramps highly depend on the orientation of the RBC in the channel cross-section at the start of the ramp. At rest, the RBC has the shape of a biconcave disk, with a rim and two dimples, schematically shown in the left simulation snapshot of Fig.~\ref{Figure_Simu}(a). The positions of the two dimples are highlighted by green and blue markers. Being subject to external forces, the RBC deforms and the dimple positions move along the RBC surface. However, after removal of the external forces, the RBC deforms back into its original shape and the rim and dimples are formed by the same part of the membrane as before the deformation, indicating a so-called shape memory of the RBC \cite{Fischer2004, Cordasco2017}. In our simulations, after the cell transitions into the slipper shape, the cell membrane tank-treads in a characteristic movement, where the two dimples follow this tank-treading in the $x$-$y$-plane, similar to previous simulations \cite{Guckenberger2018}. Two snapshots of such slipper cells are shown in the black and red boxes in Fig.~\ref{Figure_Simu}(a) for different points in time during the tank-treading. In the rightmost image of Fig.~\ref{Figure_Simu}(a), the green and blue dimples traveled a bit further than in the middle image, by rotating with the membrane in the $x$-$y$-plane. We start the downward ramps in two different simulations once when the cell is in the position shown in the middle image in Fig.~\ref{Figure_Simu}(a) and once when it is in the position shown in the right image. All other simulation parameters being equal, these configurations lead to two distinct downward transition trajectories. The $y$-coordinates of the RBCs' centers of mass during the downward transitions are compared in Fig.~\ref{Figure_Simu}(b), indicating how sensitive the system is to changes in the slipper shape at the start of the velocity ramps with respect to the membrane dimples. Although a clear correlation between the dimple positions at the start of the downward ramp and the rotating behavior or transition time could not be found, the sensitivity of the system to the exact orientation of the cell during the transition process explains the broad scattering of \mbox{$\Delta t$} in Fig.~\ref{Figure_Trans}(c).
 
\subsection*{Dynamics of slipper-shaped red blood cells}

\begin{figure*}[hbt!]
\centering
\includegraphics[width=17.145cm]{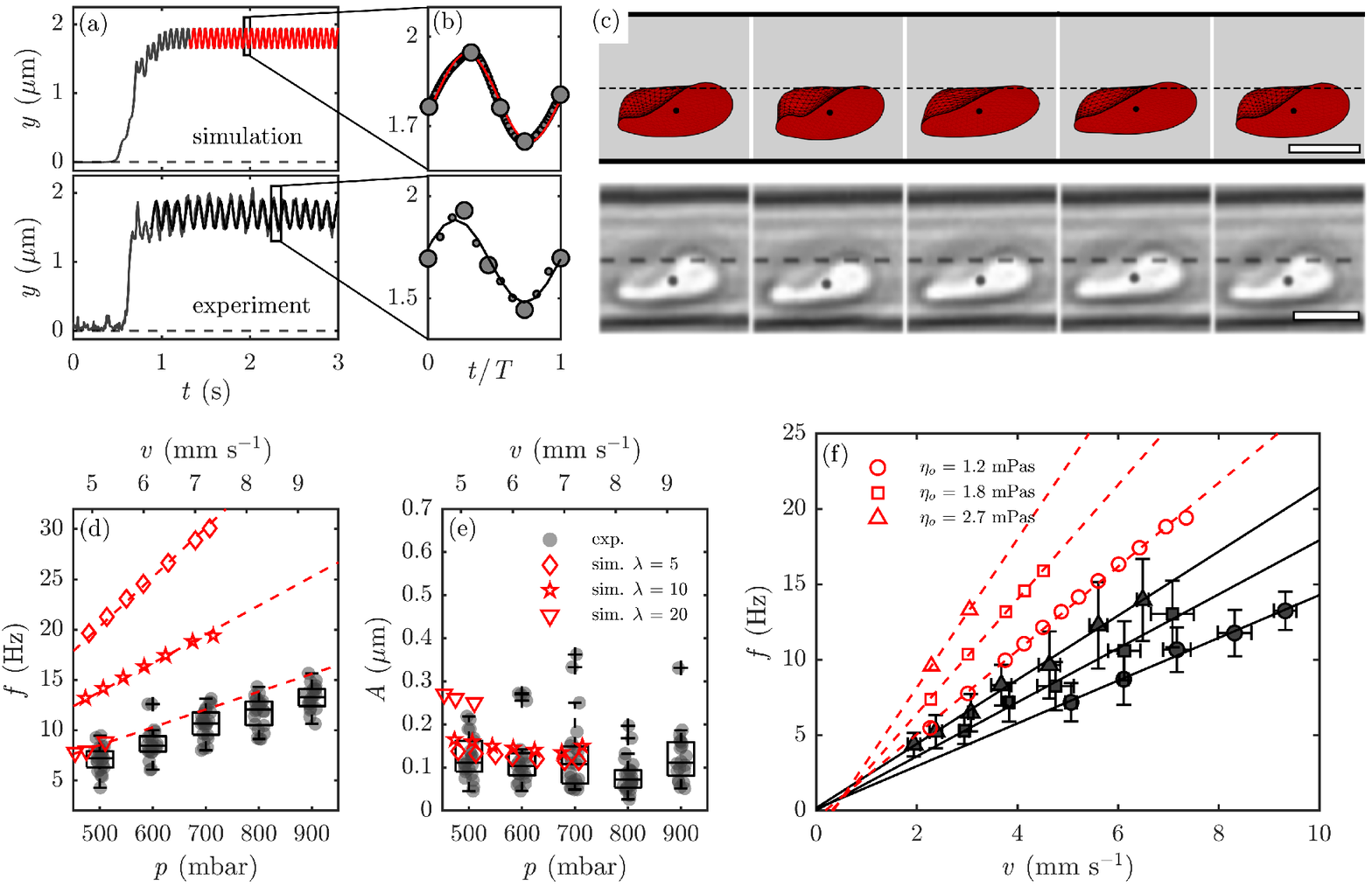}
\caption{Dynamics of slipper-shaped RBCs. Panel (a) shows the $y$-position of the cell's center of mass for a simulation (top) during a velocity ramp between \mbox{$v_{\mathrm{low}}=\unit[1]{mm\,s^{-1}}$} and \mbox{$v_{\mathrm{high}}=\unit[5.8]{mm\,s^{-1}}$}, and for an experiment (bottom) during an upward pressure ramp from \mbox{$p_{\mathrm{low}}=\unit[100]{mbar}$} to  \mbox{$p_{\mathrm{high}}=\unit[600]{mbar}$} within  \mbox{$t_{\mathrm{ramp}}=\unit[0.5]{s}$}. (b) Oscillations of the slipper-shaped cells over one oscillation period $T$, corresponding to the time windows marked by the rectangular boxes in (a). Red and black lines represent sinusoidal fits to the simulation and experimental results, respectively. (c) Representative image sequences of the slipper-shaped cells for the simulation (top) and experiment (bottom), corresponding to the points in time marked by large symbols in (b). Dashed black lines indicate the channel centerline, while the black dots represent the cell's center of mass. The scale bars represent a length of \mbox{$\unit[5]{\um}$}. The frequency and amplitude of the fits as a function of the applied pressure drop or cell velocity are shown in (d) and (e), respectively. Numerical results are plotted for different viscosity ratios \mbox{$\lambda$} as red symbols and experiments are shown as black boxplots with gray dots. Outliers are plotted using '$+$'-symbols. Panel (f) shows the oscillation frequency as a function of the cell velocity and for different outer viscosities \mbox{$\eta_\mathrm{o}$}. Black full symbols correspond to the experimental results, while red open symbols show the simulation results. Simulations are performed with a constant inner viscosity of \mbox{$\eta_\mathrm{i}=\unit[12]{mPa\,s}$}.}
\label{Figure_Osc}
\end{figure*}

As shown in Fig.~\ref{Figure_Ramp}, slipper-shaped RBCs exhibit time-dependent dynamics that lead to pronounced cell oscillations in confined channel flows. Here, we use the presented tracking technique to study these dynamics as the cells travel through the microfluidic channel. Moreover, we perform numerical simulations with different viscosity contrasts \mbox{$\lambda=\eta_\mathrm{i}/\eta_\mathrm{o}$} between the viscosity of the RBC's inner cytosol and the viscosity of the surrounding fluid. Recent simulations showed that the viscosity contrast crucially affects the flow of single vesicles and RBCs in linear shear and Poiseuille flow \cite{Danker2009, Yazdani2011, Kaoui2012, Cordasco2014, Farutin2014, Sinha2015, Mauer2018, Lehmann2020}. Additionally, the viscosity of the surrounding fluid can have an impact on the RBC shape and thus the flow properties of RBC suspensions \cite{Abkarian2008, Lanotte2016}. Therefore, we choose a broad range of \mbox{$\lambda=5$}, \mbox{$10$}, and \mbox{$20$} to study the effect of the viscosity ratio on the dynamics of slipper-shaped RBCs.

A representative cell trajectory is shown in Fig.~\ref{Figure_Osc}(a) for a simulation with \mbox{$\lambda=10$} (top) and an experiment (bottom). During this upward ramp, the RBC transitions from the croissant to the slipper shape and moves from the channel centerline to an off-centered position. Once the RBC achieves the slipper shape at  \mbox{$t\approx\unit[1]{s}$}, the cell's center of mass shows distinct oscillations around an off-centered equilibrium position of \mbox{$y\approx\unit[1.8]{\um}$} in the simulations and \mbox{$y\approx\unit[1.7]{\um}$} for the experiments. This equilibrium position does not depend on the adopted cell velocity and is not influenced by the viscosity ratio in the numerical simulations, as shown in Fig.~S7 in the Supporting Material. 

A closer look at the slipper oscillations over one oscillation period is shown in Fig.~\ref{Figure_Osc}(b) for the time windows highlighted by the rectangular boxes in (a). Additionally, snapshots of the simulation and experimental RBCs are shown in Fig.~\ref{Figure_Osc}(c) corresponding to the time frame marked in (b). While these oscillations are found in all ramp simulations for the discocyte reference shape, approximately \mbox{$\unit[50]{\%}$} of the experimentally observed slipper-shaped RBCs exhibit such pronounced oscillations. These oscillations are fitted with a sinusoidal function to extract the frequency, amplitude, and steady-state $y$-position of the cell's movement. The results are shown as red and black symbols in the lower row of Fig.~\ref{Figure_Osc} for simulations and experiments, respectively. 

\begin{figure*}[hbt!]
\centering
\includegraphics[width=17.145cm]{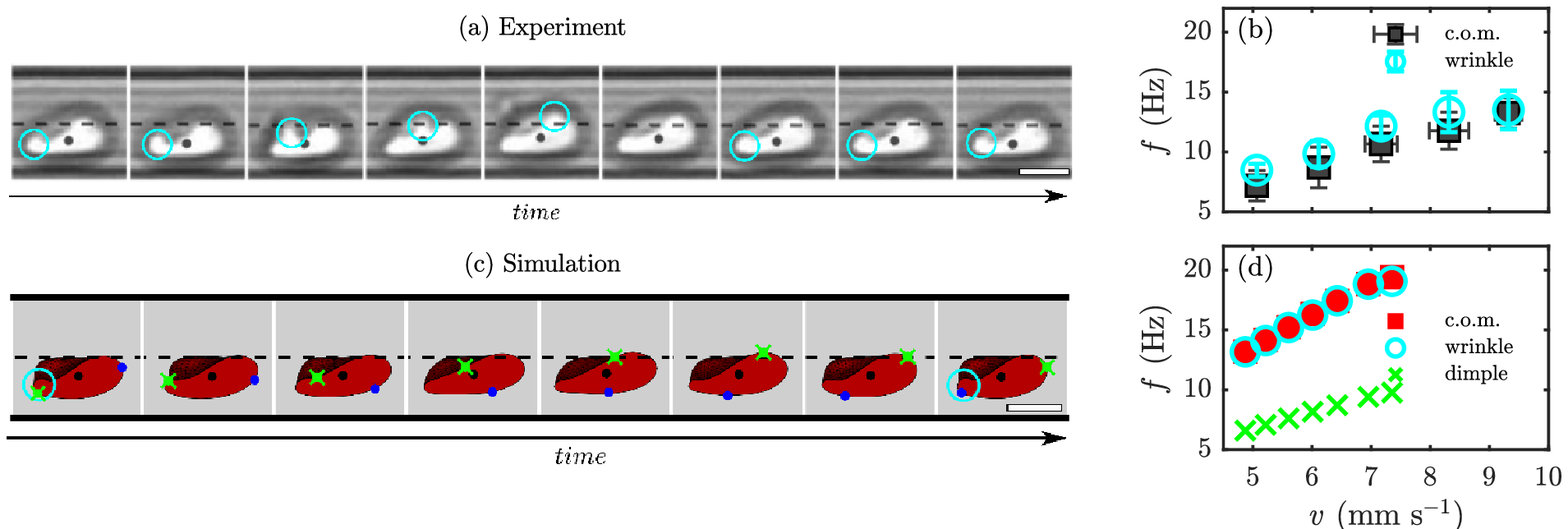}
\caption{The motion of a wrinkle on the RBC surface, marked with cyan circles in the experiment (a) and a simulation (c), at \mbox{$v\approx\unit[7]{mm\,s^{-1}}$}. The time between each consecutive image in (a) and (c) is \mbox{$\unit[10]{ms}$} and \mbox{$\unit[7.6]{ms}$}, respectively. The black dotted line indicates the channel’s centerline and dark gray dots represent the cell’s center of mass. The white scale bars correspond to a length of \mbox{$\unit[5]{\um}$}. The wrinkles in the membrane of the slipper-shaped cell periodically move along the RBC surface. The frequency of this movement is plotted in (b) and (d) as cyan symbols for the experiments and simulations, respectively, together with the frequency of oscillation of the cell’s center of mass. For the simulation in (c), the positions of the two RBC membrane dimples are highlighted with blue and green markers. The oscillation frequency of the green dimple is plotted in (d).}
\label{Figure_Membrane}
\end{figure*}

Figure~\ref{Figure_Osc}(d) shows the oscillation frequency as a function of the applied pressure drop or cell velocity. For both simulation and experiment, the oscillation frequency increases linearly with increasing cell velocity. Further, increasing the viscosity ratio in the simulations results in a decrease of the oscillation frequency, which is connected to the tank-treading frequency of the membrane. Similarly, earlier investigations in shear flow have shown that the tank-treading frequency of RBCs is influenced by the viscosity contrast \cite{Fischer2007, Abkarian2007, Fedosov2010, Yazdani2011a}. In our study, simulations with \mbox{$\lambda=20$} fit the experimental observations best, whereas simulations with a lower viscosity contrast deviate from the experiments by a factor of approximately 2.8 and 1.9 for \mbox{$\lambda=5$} and \mbox{$\lambda=10$}, respectively. While the frequency increases linearly with the cell velocity, the amplitude of the slipper oscillations is not systematically affected by the applied pressure drop or velocity, as shown in Fig.~\ref{Figure_Osc}(e). For the experiments, the average amplitude is roughly \mbox{$\unit[0.12]{\um}$}. For the simulations with \mbox{$\lambda=5$} and \mbox{$\lambda=10$}, we find a similar mean amplitude of \mbox{$\unit[0.12]{\um}$} and \mbox{$\unit[0.14]{\um}$}, respectively, in good agreement with the experimental observations. However, \mbox{$\lambda=20$} yields a higher average amplitude of \mbox{$\unit[0.26]{\um}$}. Note that in the numerical simulations with \mbox{$\lambda=20$}, slipper-shaped RBCs exhibit different behavior in two consecutive regimes. First, the cells show the regular tank-treading slipper movement with dimples following the membrane rotation in the $x$-$y$-plane. However, after a few oscillation cycles, the cell rotates such that the dimples are located at the side of the tank-treading slipper. This reorientation of the membrane influences the oscillation frequency and amplitude of the cell's center of mass and is also found for \mbox{$\lambda=10$}. However, for \mbox{$\lambda=10$} the first regime is stable for a longer time, as shown in Fig.~S8 in the Supporting Material. In our simulations, a lower viscosity contrast and larger cell velocities seem to stabilize the initial regular slipper oscillations. The results in Fig.~\ref{Figure_Osc} show the data obtained from fitting the oscillations in this first stable regime. After the reorientation, the slipper oscillations in $y$-direction vanish almost completely. This effect might also be reflected in the experiments. We hypothesize that the fact that pronounced oscillations are found in only \mbox{$50\%$} of all observed slippers is linked to a distribution of the cytosol viscosity within the RBC population. Hence, we do not observe slipper oscillations in the channel's $y$-direction for cells with a larger \mbox{$\lambda$}, while cells with a small viscosity contrast show distinct oscillations. 

Moreover, we explore the effect of different RBC elastic reference shapes for the shear contribution on the dynamics of slipper-shaped cells in the numerical simulations. Here, simulations are performed with the typical discocyte as reference shape as well as two oblate spheroids with aspect ratio \mbox{$\tau=0.9$} or \mbox{$\tau=0.98$}. As shown in Fig.~S9 in the Supporting Material, we find a similar oscillation frequency for the oblate reference shape with \mbox{$\tau=0.9$} as for the discocyte reference shape. For \mbox{$\tau=0.98$}, the slipper oscillations show a higher frequency. Further, both oblate spheroids lead to much smaller oscillation amplitudes than the discocyte, in disagreement with the experimental observations. Also, variations of the flat bending reference shape, the shear modulus \mbox{$\kappa_\mathrm{S}$}, and the bending modulus \mbox{$\kappa_\mathrm{B}$} within the range of physiological values for RBCs have negligible impact on the simulation results, as shown in Fig.~S10 in the Supporting Material.

While the viscosity contrast can be arbitrarily adjusted in the numerical simulation, the viscosity of the RBC's cytosol in the experiments cannot be determined precisely. To study the effect of the viscosity contrast on the slipper dynamics experimentally, we perform RBC tracking experiments in different dextran solutions as the surrounding fluid, resulting in outer viscosities in the range between \mbox{$1.2\leq \eta_\mathrm{o}\leq\unit[2.7]{mPa\,s}$}, as shown in Fig.~\ref{Figure_Osc}(f). The $x$-axis of Fig.~\ref{Figure_Osc}(f) corresponds to the measured cell velocity. Complementary simulations are performed at the same outer viscosity values and a constant inner viscosity of \mbox{$\eta_\mathrm{i}=\unit[12]{mPa\,s}$}. Both experiments and simulations show an increase of the oscillation frequency with increasing outer viscosity as well as a linear dependency on the RBC velocity. Further, the simulations deviate from the experimental observations by a constant factor of roughly 1.9, independent of the viscosity of the surrounding fluid. These deviations might arise due to the lack of the membrane viscosity as an additional parameter in the present numerical simulations besides \mbox{$\eta_\mathrm{i}$} and \mbox{$\eta_\mathrm{o}$}. Recent simulations of capsules and RBCs suggest that the incorporation of the membrane dissipation through a shear-thinning membrane viscosity affects the tank-treading dynamics of RBCs and yields good agreement with experimental observations in simple shear flow \cite{Guglietta2020b, Matteoli2021}. Our simulations also show that although increasing the viscosity contrast has similar effects as including or increasing the viscosity of the membrane, these two parameters are not qualitatively identical \cite{Guglietta2020a, Li2021}. Moreover, we experimentally observe that for an outer viscosity of \mbox{$\eta_\mathrm{o}=\unit[1.2]{mPa\,s}$}, slipper-shaped RBCs predominantly appear above a cell velocity of roughly \mbox{$v=\unit[5]{mm\,s^{-1}}$}. However, increasing the outer viscosity to \mbox{$\eta_\mathrm{o}=\unit[2.7]{mPa\,s}$}, we also find oscillating slipper-shaped RBCs at lower cell velocities of \mbox{$v\approx\unit[2]{mm\,s^{-1}}$}, as shown in Fig.~\ref{Figure_Osc}(f). This suggests that the external shear stress on the cell influences the RBC shape \cite{Abkarian2008}, which indicates a dependence of the RBC shape phase diagram not only on the intrinsic mechanical RBC properties but also on the surrounding fluid. 

The origin of the observed oscillations of slipper-shaped cells in Fig.~\ref{Figure_Osc} is a tank-treading movement of the RBCs that has been investigated in rheoscopic experiments before \cite{Fischer1978, Tran-Son-Tay1984, Fischer2007}. In the microfluidic channel, this rotating movement of the RBC membrane in combination with the confinement of the cell and the strong shear gradient in the narrow channel leads to the distinct oscillations of the cell's center of mass in the $y$-direction. In some experiments where cells show pronounced oscillations, we observe the formation of a wrinkle in the RBC membrane that travels over the RBC surface (Movie~S5 in the Supporting Material). Cyan circles in Fig.~\ref{Figure_Membrane}(a) highlight the position of such a wrinkle as it travels along the RBC surface during a recorded image sequence. The frequency of this movement is similar to the oscillation frequency of the cell's center-of-mass, as shown in Fig.~\ref{Figure_Membrane}(b). In the numerical simulation, similar wrinkles are also observed during certain stages of the oscillation, as shown at the cell's tail in the first and last images of Fig.~\ref{Figure_Membrane}(c) for \mbox{$\lambda=10$} (Movie~S6 in the Supporting Material). The frequency with which this wrinkle appears at the tail of the slipper is plotted in Fig.~\ref{Figure_Membrane}(d) and coincides with the oscillation frequency of the cell's center of mass, similar to the experimental observations. 

Furthermore, we can mark specific spots, \textit{e.g.}, the dimples in the initial discocyte RBC shape, on the RBC membrane in the numerical simulations and track their positions over time. The movement of these marked spots ultimately corresponds to the tank-treading movement of the cell. The tank-treading frequency of one dimple, marked by the green marker in Fig.~\ref{Figure_Membrane}(c), is exactly half the frequency of the center of mass movement as well as the frequency of the wrinkle, as shown in Fig.~\ref{Figure_Membrane}(d). In the first image of the sequence in Fig.~\ref{Figure_Membrane}(c), the wrinkle emerges at a position before the green dimple, while in the last image, another wrinkle forms before the blue dimple. Hence, two wrinkles are formed at an interval of half the cell's circumference in the $x$-$y$-plane that travel over the RBC membrane at twice the tank-treading frequency. In the simulations, we observe that wrinkles are more pronounced for higher viscosity ratios. Experimentally, roughly \mbox{$5\%$} of all slipper-shaped RBCs that show pronounced oscillations exhibit such wrinkles, independent of the applied pressure drop. We hypothesize that the appearance of moving wrinkles in the RBC membrane is related to the distribution of cell age and therefore the mechanical RBC properties. Detailed investigations of these peculiar dynamics of RBCs in confined microchannel flow will therefore be performed for age-separated cell fractions in future studies.

\section*{Conclusion}
The flow behavior and distinct shapes of RBCs play a pivotal role in the microvascular blood flow. In microfluidic studies with channel dimensions similar to the RBC diameter, the symmetric croissant in the channel center and the non-symmetric slipper closer to the vessel walls are the dominant RBC shapes, depending on the cell velocity as determined by the applied pressure drop. In this study, we analyzed the dynamics of single RBCs in confined, time-dependent flows using a direct comparison of microfluidic experiments and numerical simulations. Therefore, we introduce a customized, feedback-controlled tracking technique that enables us to investigate the shape evolution of cells while traveling through the microchannel. Employing increasing and decreasing pressure ramps, we find that the transition from the centered croissant shape to the off-centered slipper shape is with a mean transition time of \mbox{$ \overline{\Delta t}_{\mathrm{up}}=\unit[0.36]{s}$} much faster than the opposite shape transition with \mbox{$ \overline{\Delta t}_{\mathrm{down}}=\unit[1.6]{s}$}. During the latter shape transition, we observe rich rotation and tumbling events that prolong the shape change. Complementary numerical simulations reveal that such events highly depend on the position of the cell's characteristic dimples of the biconcave RBC shape at the start of the transition process. 

Moreover, the presented tracking method allows us to experimentally probe the temporal dynamics of slipper-shaped RBCs along the flow direction. Here, we report a unique RBC movement that is characterized by distinct lateral oscillations of slipper-shaped RBCs towards the channel sidewalls in $y$-direction. Our numerical simulations reveal that the origin of these characteristics is a tank-treading of the RBC membrane in combination with the strong confinement in the narrow microchannel. The frequency of these oscillations increases with the cell velocity and with the viscosity of the surrounding fluid. In contrast, the oscillation amplitude is not affected by the cell velocity or the applied pressure drop.

The analysis of single RBCs in flow provides a versatile tool to identify and examine the influence of pathological changes of RBC mechanical properties on their morphological dynamics, especially for cells with impaired deformability. Furthermore, the presented technique can be used in future studies to probe the influence of plasma composition or drug administration on RBCs dynamics, which could potentially provide a toolbox for lab-on-a-chip medical devices. From an experimental perspective, the presented approach can be further used to study the dynamics and interactions of other objects, \textit{e.g.}, cells, bacteria, algae, in microscale flows. In the context of hematology, the consequent next step will be to employ the comoving tracking to provide experimental details about the hydrodynamic interactions between multiple RBCs in capillary flows. Moreover, plasma proteins, such as fibrinogen, can be added to the surrounding fluid to tune the interactions between individual RBCs and to probe the emerging dynamics with respect to the time scale of the time-dependent flow, thus, enabling a better understanding of microvascular blood flow.

\section*{Author Contributions}
SMR conceptualized the study, conducted the experiments, analyzed and interpreted the data, and wrote the manuscript. KG performed the numerical simulations, analyzed and interpreted the data, and wrote the manuscript. FMM designed the feedback-controlled tracking algorithm and revised the manuscript. TJ interpreted the data and revised the manuscript. SG and CW supervised the study and revised the manuscript.

\section*{Acknowledgments}
This work was supported by the Deutsche Forschungsgemeinschaft DFG in the framework of the research unit FOR 2688 `Instabilities, Bifurcations and Migration in Pulsatile Flows' WA 1336/13-1 and GE 2214/2-1. SMR gratefully acknowledges the help of Mr. Mohammed Nouaman from Saarland University regarding the phase diagram for stable RBC shapes. KG thanks the Studienstiftung des deutschen Volkes for financial support. KG and SG acknowledge the work of Dr. Achim Guckenberger, who implemented the boundary integral method.

\bibliography{main}


\end{document}